\documentclass[twocolumn,showpacs,preprintnumbers,amsmath,amssymb]{revtex4}

\usepackage{graphicx}
\usepackage{dcolumn}
\usepackage{bm}

\begin{document}

\title{Dyadic Cantor set and its kinetic and stochastic counterpart}%

\author{M. K. Hassan$^{1}$,  N. I. Pavel$^{1}$, R. K. Pandit$^{1}$ and J. Kurths$^{2}$
}%
\date{\today}%

\affiliation{
$1$ University of Dhaka, Department of Physics, Theoretical Physics Group, Dhaka 1000, Bangladesh \\
$2$ Potsdam Institute for Climate Impact Research, Telegrafenberg A31, 14473 Potsdam, Germany
}

\begin{abstract}

Firstly, we propose and investigate a dyadic Cantor set (DCS) and its kinetic counterpart where a generator 
divides an interval into two equal parts and removes one with probability $(1-p)$. The generator is then applied 
at each step to all the existing intervals in the case of DCS and to only one interval, picked with probability according 
to interval size, in the case of kinetic DCS. Secondly, we propose a stochastic DCS in which, unlike the kinetic DCS, the generator 
divides an interval randomly instead of equally into two parts. Finally, the models are solved analytically; an exact expression for 
fractal dimension in each case is presented and the relationship between fractal dimension and the corresponding conserved 
quantity is pointed out. Besides, we show that the interval size distribution function in both variants of DCS exhibits 
dynamic scaling and we verify it numerically using the idea of data-collapse.
\end{abstract}


\maketitle


\section{Introduction}

The world we live in is not restricted to  Euclidean geometry like lines, squares, rectangles, circles, semi-circles, 
spheres, etc only. 
Instead, there are curves twisted so wildly that they occupy  plane not a line, there are surfaces 
folded so badly that they occupy space not a plane, there are objects so stringy or ramified 
that their constituents are distributed scarcely not uniformly. 
Indeed, most of the natural objects we see around us are so complex in shape 
and structure that Euclidean geometry is not sufficient to describe them. Many of these objects used to be described as
geometrically chaotic, since they are not just merely complex but often contain different
degrees of complexity. In 1975 Mandelbrot introduced the idea of fractal to describe these geometrically chaotic objects 
and it has revolutionized the notion of geometry forever. It helps us appreciate the fact that there exists some kind 
of order even in seemingly disordered and apparently bewildering objects \cite{ref.mandelbrot1,ref.mandelbrot2}.
Prior to the inception of the idea of fractal, geometry remained one of the main branches of mathematics. However,
soon after its inception, it has 
attracted mathematicians, physicists, and engineers all alike and hence generated a widespread interest in subjects like 
physics, chemistry, biology, earth science, economics, etc \cite{ref.vicsek,ref.korvin,ref.peters}.
The exact definition of fractal remains  elusive even after more than thirty five years.
This is partly because Mandelbrot himself was somewhat reluctant to confine it within the 
boundaries of a mere definition. He nevertheless proposed that fractal is
a geometric object with irregular shape made of parts similar to the whole, in some sense. 


The idea of the Hausdorff-Besicovitch dimension, however, plays a pivotal role in defining fractal.
Consider that the measure $H_d$ describes the size of the
set of points that constitute an Euclidean object. We can quantify the size of the measure $H_d$ by using a $d$-dimensional
hyper-cube of linear size $\delta$ as an yardstick. We can write the measure $H_d$ in terms of the
 number $N(\delta)$ needed to cover the object as
\begin{equation}
\label{eq:measure}
H_d=\sum \delta^d=N(\delta)\delta^d,
\end{equation}
where $\delta^d$ is the test function \cite{ref.feder}.  It can be easily shown that $N(\delta)$ for Euclidean objects 
always satisfies 
\begin{equation}
\label{eq:gdim5}
N(\delta/n)=n^{d_E} N(\delta)  \ \ \ \ {\rm with} \ \ \ \   d_E=1,2,3,
\end{equation}
and hence $N(\delta)$ is a homogeneous function.
One can explicitly prove that only a power-law solution for $N(\delta)$ can satisfy Eq. (\ref{eq:gdim5}) \cite{ref.newman}. Indeed, 
it is easy to check that
\begin{equation}
\label{eq:gdim6}
N(\delta)\sim \delta^{-d_E},
\end{equation}
is a solution of Eq. (\ref{eq:gdim5}). It implies the dimension $d_E$ of an object can be defined as the slope of the plot of 
$N(\delta)$ vs $\delta$ in the $\log-\log$ scale. 
Obviously, the slope $d_E$ of such a plot for Euclidean objects will always be an integer quantity.  
However, there exists another class of natural or man-made objects for which the slope of the same plot 
may assume a non-integer value which we typically denote by $d_f$. We can therefore generalize the solution for $N(\delta)$ 
upon replacing $d_E$ by an unconstrained exponent $D$. Using it in Eq. (\ref{eq:measure}) we find that there exists a
critical dimension $d=D$,  known as  the Hausdorff-Besicovitch (H-B) dimension, for which the measure $M_d$ neither vanishes
nor diverges as $\delta \rightarrow 0$ \cite{ref.feder}. Thus, an object is fractal if the H-B dimension $D=d_f$ assumes
a non-integer value and at the same time it is less than the dimension of the space where the object is being embedded.

One of the best known text-book example of fractal is the triadic Cantor set. It
starts with an initiator, typically an interval of unit size, and a generator that divides 
it into three equal parts and remove the middle third. The generator is then 
applied at each successive step 
to all the available smaller intervals over and over again.  The resulting set has a non-integer H-B dimension 
$d_f=\ln 2/\ln 3$
with numerical value less than that of the space $d=1$ where the set is embedded \cite{ref.feder}. 
Besides its pedagogical importance, the Cantor set problem has also been of theoretical and practical interest \cite{ref.sears,ref.hatano,ref.redner,ref.esaki}.
However, as far as natural fractals are concerned, the Cantor set lacks at least in 
two ways. Firstly, it does not appear through evolution in time,
although fractals in nature do so. Secondly, it is not governed by any sort of randomness 
throughout its construction process, while natural fractals always occur through some kind of evolution 
accompanied by some randomness. Note that nature likes freedom not determinism.
Indeed, it is well understood that in our world almost nothing is stationary
or strictly deterministic. Most natural objects we see around us are seemingly complex in character albeit they are governed
by simple rules. Note that it is the simple rules when repeated over and over again that make the resulting system 
look  mighty complex in the end.

Motivated by the importance and impact of the Cantor set, we investigate here a couple of its interesting variants
in which probability, time and randomness are incorporated in a logical progression. 
To this end, we first propose dyadic Cantor set (DCS) which is simpler than the much known triadic Cantor set (TCS)
since dividing into two is undeniably simpler than dividing into three in any sense.  We then introduce the kinetic 
counterpart of the DCS by applying the generator to only one interval at each step by picking it preferentially with 
respect to interval size. 
The sequential application of the generator immediately 
introduces time as one parameter in the problem and hence it helps us to understand its impact on the resulting fractal
and various other observable quantities. Finally, we introduce spatial randomness by modifying
 the generator such that it divides an interval randomly into two parts instead of dividing into two equal parts and apply 
it sequentially like kinetic DCS. Consequently, it now incorporates 
both time and spatial randomness in the system making it a stochastic counterpart of the DCS. 
Each variants of the DCS 
are solved analytically to obtain the fractal dimension and to show self-similarity. Analytic results, especially the
self-similar properties, are verified numerically by invoking the idea of data collapse.

The remainder of this article is organized as follows. In section II, we propose the dyadic Cantor set
and discuss various properties. In section III, 
we propose the kinetic counterpart of the dyadic Cantor set and solve it exactly to obtain the fractal dimension
and various other properties. In section IV, we investigate the stochastic counterpart
of the dyadic Cantor set and solve it analytically. We also propose its exact algorithm to solve it by numerical simulation to 
verify our analytical results. We then 
revisit the kinetic DCS problem in section V with an aim to check whether it also exhibits dynamic scaling or not. 
Finally, in section VI, we give a general discussion and summary of the work.

\section{Dyadic Cantor set (DCS)}

In this section, we first define the dyadic Cantor set (DCS) and then investigate its various aspects.
It starts with an initiator which is typically an interval  of unit length $[0,1]$. The generator then divides the initiator into 
two equal parts 
and deletes one, say the right half, with probability $(1-p)$. After step one, the system on average will have  $(1+p)$ number of sub-intervals 
of size $1/2$, since the right half interval remains there in step one with probability $p$.
In the next step, the generator is applied to each of
the $(1+p)$ available  sub-intervals to divide them into two equal parts and remove the right half from each of the
$1+p$ intervals with probability $(1-p)$. The system will then have on average $(1+p)^2$ number of intervals of size $1/4$ as $(1-p)(1+p)$ number of
intervals of size $1/4$ are removed on  average.   
The process is then continued over and over again by applying the generator on all the available intervals at each successive
step recursively.

Finding the fractal dimension of the DCS problem is as simple as its definition. 
According to the construction 
of the DCS process, there are $N=(1+p)^n$ intervals in the $n$th generation and they are of size $\delta=2^{-n}$.
 The most convenient yard-stick to measure the size of the set in the
$n$th step is the mean interval size $\delta=2^{-n}$. Expressing $N$ in favour of $\delta$ by using 
$\delta=2^{-n}$, we find 
\begin{equation}
\label{eq:fractal}
N(\delta)\sim \delta^{-d_f},
\end{equation}
where $d_f={{\ln (1+p)}\over{\ln 2}}$. 
Note that the exponent $d_f$ is non-integer for all $0<p<1$ and at the same time 
it is less than the dimension of the space $d=1$, where the set is embedded. It is therefore the fractal dimension of the resulting
DCS which does not fill up the unit interval continuously to constitute a line. Unlike the triadic Cantor set where the 
Cantor dusts are distributed in a strictly self-similar 
fashion, the Cantor dusts in the dyadic Cantor set are distributed in a random fashion, yet it is self-similar
in the statistical sense only.

One of the most interesting aspects of the triadic Cantor set is worth mentioning here. The intervals which
are removed from the triadic Cantor set are of size $1/3, 2/9, 4/27, ...$ etc and if we add them up we get 
\begin{equation}
{{1}\over{3}}\sum_{n=0}^\infty\Big ({{2}\over{3}}\Big )^n=1.
\end{equation}
This is the size of the initiator and hence it means that there is nothing left in the triadic Cantor set, since 
the sum of the sizes that
are removed equals the size of the initiator. Interestingly, we find that a similar argument holds for dyadic Cantor set too. For instance, on the average in step one the amount of size removed is ${{1-p}\over{2}}$, in step two the 
total amount of size removed is ${{(1-p)(1+p)}\over{4}}$, in step three it is ${{(1-p)(1+p)^2}\over{8}}$, in step four it is ${{(1-p)(1+p)^3}\over{16}}$ 
and so on. If we add these intervals we have
\begin{equation}
{{(1-p)}\over{2}}\sum_{n=0}^\infty \Big ({{1+p}\over{2}}\Big )^n
=1
\end{equation}
which is again the size of the initiator. It means, like in the triadic Cantor set, there is hardly anything left in the DCS. However, 
we will show later that there are still tons of members in the set.

\section{Kinetic dyadic Cantor set}

We now ask: What happens if the generator of the DCS is applied to only one interval 
at each step instead of 
applying it to all the available intervals in each successive step?  
Clearly, the spatial distribution of the remaining intervals will be very different from the one created by the DCS problem. 
Yet the question is: Will the number $N$ needed to cover the set by an yardstick, say the mean interval 
size $\delta$, still exhibit a power-law against $\delta$? If yes, will the exponent {\it vis-a-vis} the 
fractal dimension be the same as that of the DCS? To find a definite answer to these questions, it is important to appreciate 
the fact that 
after step one and beyond the system will have intervals of different sizes and hence it raises a further
question: How do we choose one interval when the system has intervals of different sizes? To this end, we choose
the most generic case whereby an interval is picked preferentially
 with respect to their sizes.

The algorithm of the kinetic DCS problem can be defined as follows. Like DCS it also starts with an initiator of a unit interval 
$[0,1]$. In the first step the generator divides the initiator into two sub-intervals of equal size
and removes one of the part, say the right half, with probability $(1-p)$. 
There are now $(1+p)$ intervals each of size $1/2$. In the next
step we generate a random number $R$ from the open interval $(0,1)$ and find which of the $(1+p)$ 
sub-intervals contains this number $R$ in order to ensure that the interval is picked preferentially according to their sizes. 
If $R$ is found within the interval $[0,{{1}\over{2}}]$ then we pick it; else we pick the right interval
if it has not already been removed. Say, the left interval contains $R$ and hence we pick that and 
the generator is then applied onto it only. 
In any case, time is increased by one unit even if $R$ is found within
the interval that has been removed in which case the generator is not applied at all.

In order to study the kinetic DCS problem analytically, we use the binary fragmentation equation  \cite{kn.ben_naim_source,ref.ziff}
\begin{eqnarray}
\label{eq:binary_frag}
 \frac{\partial c(x,t)}{\partial t}&=& - c(x,t)\int_{0}^{x}dyF(y,x-y)\\ \nonumber &+& 2\int_{x}^{\infty}dyF(x,y-x)c(y,t),
\end{eqnarray}
where $c(x,t)dx$ is the number of intervals within the size range $x$ and $x+dx$ at time $t$ and $F(x,y)$ is the fragmentation kernel that specifies the 
rate and the rules of the fragmentation process. The first term on the right hand side of the above equation 
describes the loss of interval of size $x$ due to its breakup into two smaller intervals, while the second term describes the gain of interval of size $x$ due to 
breakup of an interval of size $y>x$ into two smaller pieces ensuring that one of the two smaller intervals is of size $x$. 
The factor $2$ in the gain
 term implies that at each time step two intervals are produced out of one interval.  
However, in the context of the present model we need to replace the factor $2$ of Eq. (\ref{eq:binary_frag}) by $(1+p)$
 to ensure that on the average at each breaking event $(1+p)$ number of new 
intervals are produced. Next, we need to choose
 the following kernel 
\begin{equation}
\label{eq:kernel_delta}
 F(x,y)=(x+y)\delta(x-y),
\end{equation}
to ensure that at each breaking event an interval is picked preferentially with respect to interval size and divide 
it into two equal pieces. Using the two facts in Eq. (\ref{eq:binary_frag}) we obtain 
\begin{equation}
\label{eq:KDCS}
\frac{\partial c(x,t)}{\partial t}=-\frac{x}{2}c(x,t)+2x(1+p)c(2x,t),
\end{equation}
which is the required  fitting equation that describes the kinetic DCS problem.

We now find it convenient to incorporate the definition of the moment
\begin{equation}
M_n(t)=\int_{0}^{\infty}x^n c(x,t)dx,
\end{equation}
 in Eq. (\ref{eq:KDCS}) 
which gives the rate equation for $M_n(t)$ that reads as
\begin{equation}
\label{eq:kinetic_moment}
\frac{dM_n(t)}{d t}=-\Big [\frac{1}{2}-\frac{(1+p)}{2^{(n+1)}}\Big ]M_{n+1}(t).
\end{equation} 
This equation reveals that there exists a value of $n=n^*$ for which the moment $M_{n^*}$ is a conserved quantity. The value of $n^*$
can be found simply by finding the root of the 
following equation
\begin{equation}
 \frac{1}{2}-\frac{(1+p)}{2^{(n^*+1)}}=0.
\end{equation}
Solving it we obtain $n^*= \frac{\ln(1+p)}{\ln 2}$ implying that the moment $M_{\frac{\ln(1+p)}{\ln 2}}(t)$ 
is a conserved quantity. To verify it numerically  we denote the sizes of all the surviving intervals at the $j$th  step as 
$x_1,x_2,x_3,..., x_{N_j}$ starting from the left most till the right most within $[0,1]$
 in order to avoid the use of $c(x,t)$ in the definition of the moment. 
The $n^*$th moment then is
\begin{equation}
M_{n^*}=\sum_{i=1}^{N_j}x_i^{n^*},
\end{equation} 
which is indeed a conserved quantity according to Fig. (1). However, it is worth mentioning that the numerical value of the
conserved quantity $M_{\frac{\ln(1+p)}{\ln 2}}(t)$ is not the same at every independent realization. The question remains: 
Why the index of the moment $n^*= \frac{\ln(1+p)}{\ln 2}$ is so special? 
To find an answer to this question we invoke the  idea of fractal analysis below.

It is expected that the kinetic DCS problem too, like the simple DCS, will generate fractal in the long time limit and hence 
it must possess self-similarity as it is
an essential property of fractal. That is, the various moments of $c(x,t)$, which correspond to 
observable quantities, should exhibit a power-law relation with time,
since dimensional functions are always power-law monomial in character. We therefore 
can write a tentative solution for the $n$th moment as  
\begin{equation}
\label{eq:ten_moment}
M_n(t) \sim A(n)t^{\alpha(n)}.
\end{equation}
If we insist that it must obey the conservation law, $M_{{{\ln (1+p)}\over{\ln 2}}}=const.$ then the exponent $\alpha(n)$ must satisfy 
$\alpha(n^*)=0$. Substituting Eq. (\ref{eq:ten_moment}) in Eq. (\ref{eq:kinetic_moment}) yield the following recursion relation
\begin{equation}
\alpha(n+1)=\alpha(n)-1.
\end{equation}
Iterating it subject to the condition that $\alpha(n^*)=0$ gives
\begin{equation}
\alpha(n)=-\Big (n-{{\ln (1+p)}\over{\ln 2}}\Big ).
\end{equation}
We therefore now have an explicit asymptotic solution for the $n$th moment  
\begin{equation}
\label{eq:moment_e}
M_n(t) \sim t^{-\Big(n-{{\ln (1+p)}\over{\ln 2}} \Big )}.
\end{equation}
It means that the number of intervals $N(t)=M_0(t)$ grows with time as
\begin{equation}
\label{eq:number}
N(t)\sim t^{{{\ln (1+p)}\over{\ln 2}}},
\end{equation}
revealing that there exists a non-trivial relation between the number of intervals $N$ and the time $t$. 
On the other hand we find that the mean interval size $\delta=M_1(t)/M_0(t)$ decreases with time as
\begin{equation}
\label{eq:mean_krdcs}
\delta(t)  \sim t^{-\gamma},
\end{equation}
where the exponent $\gamma = 1$.

\begin{figure}
\includegraphics[width=5.50cm,height=8.5cm,clip=true,angle=270]{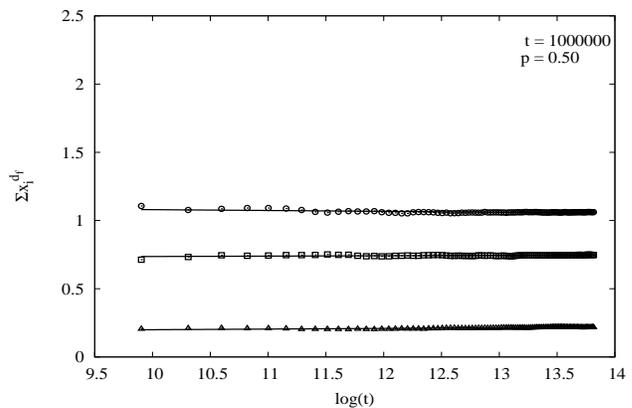}
\caption{The three horizontal lines for three independent realizations for the same $p$ value show that the $d_f$th moment $M_{d_f}$
of the remaining intervals,
where $d_f=\ln (1+p)/\ln 2$, is a conserved quantity as suggested by Eq. (\ref{eq:moment_e}). However, note that
the numerical value
of $M_{d_f}$ assumes different numerical value at each independent realization. 
}
\label{fig1}
\end{figure}

To verify Eq. (\ref{eq:number}) numerically we plot $\ln N$ vs $\ln t$ in Fig (2) 
and find a straight line with slope exactly  equal to $\ln (1+p)/\ln 2$ as expected according to Eq. (\ref{eq:number}). 
 In fractal analysis, one usually seeks for a power-law relation between the number $N(\delta)$ 
and a suitable yard-stick size $\delta$.  The mean interval size $\delta$ is definitely 
the best choice for the yard-stick. To find the required relation we therefore 
eliminate $t$ from Eq. (\ref{eq:number}) in favour of $\delta$ given by Eq. (\ref{eq:mean_krdcs}) 
 and we immediately find that  $N$ exhibits the same power-law $N\sim \delta^{-d_f}$
as we found in the case of DCS including the exponent $d_f ={{\ln (1+p)}\over{\ln 2}}$. It implies
that  the value of the index $n^*$ of the conserved quantity is actually the fractal dimension.
One could not help but check about the $d_f$th moment in the DCS case too. In this case the $d_f$th moment in the $n$th step is  
\begin{equation}
M_{d_f}=\sum_{n=1}^{(1+p)^n} x_i^{d_f}=1,
\end{equation}
and hence it is indeed a conserved quantity since all the intervals are of same size regardless of the $n$ value.   
The same is also true for the triadic Cantor set and one can easily verify it by setting $x_{i}=3^{-n}$, $N=2^n$ and $d_f=\ln 2/\ln 3$.
This is surprising in the sense that on one hand the sum of all the intervals which are removed is
equal to the size of the initiator revealing there is nothing left in the set.
On the other hand, the $d_f$th moment $M_{d_f}$ in all cases is a conserved quantity revealing that the system 
still has tons of intervals. The connections between the fractal dimension and the conserved
quantity was first reported by  Ben-Naim and Krapivsky in the context of 
the stochastic Cantor set \cite{ref.fractal}. Later, we found such connections to be true in many different varients of fragmentation
and aggregation processes \cite{ref.fractalCDA,ref.fractalself_repl,ref.hassanRogers1,ref.hassanRogers2,ref.hassan1,ref.hassan2}. 

\begin{figure}
\includegraphics[width=5.50cm,height=8.5cm,clip=true,angle=270]{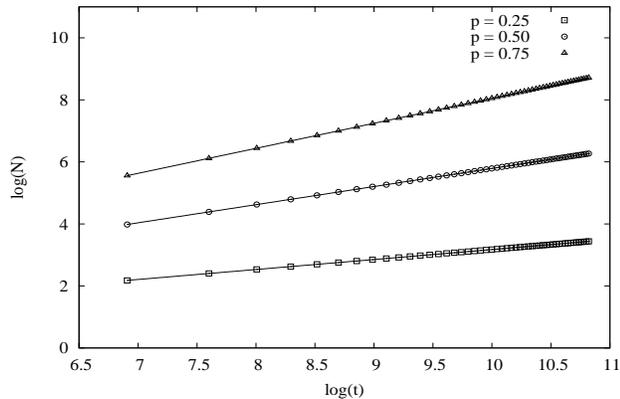}
\caption{ ${\rm Log}-\log$  plot of the interval number $N$ vs time $t$ are drawn for the kinetic DCS using three different $p$ values.
 The lines have slope
equal to fractal dimension $d_f=\ln(1+p)/\ln 2$, revealing that $N(t) \sim t^{d_f}$ as predicted by Eq. (\ref{eq:number}).
}
\label{fig2}
\end{figure}

\section{Stochastic dyadic Cantor set}

It is true that fractal in the DCS has some form of spatial randomness albeit the intervals at any stage of the construction process are of equal size. 
In contrast, intervals in the kinetic DCS are distributed not only randomly but they are also of different size at any given
stage. Yet, we find that they share the same fractal dimension owing to the same deterministic rule in the definition 
of the generator. Nature do not like determinism rather it likes to enjoy some form of freedom in the selection process. 
Freedom in the present context means the liberty to divide an interval randomly.  
We therefore ask: What if we use a generator that divides an interval randomly into two smaller intervals instead of dividing
it into two equal intervals? To this end we propose the following stochastic dyadic Cantor. We start the process with an 
initiator of unit interval $[0,1]$ as before but unlike the previous cases the generator here divides an interval 
randomly into two pieces and removes one with probability $(1-p)$. The algorithm
of this model is exactly the same as the one for kinetic DCS except the fact that 
the generator always divides an interval randomly into two parts instead of dividing into two equal parts.

\begin{figure}
\includegraphics[width=5.50cm,height=8.5cm,clip=true,angle=-90]{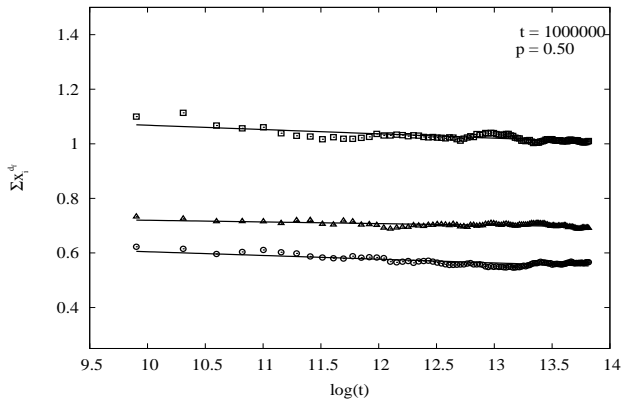}
\caption{Plot of the $p$th moment $M_p(t)$  of the remaining intervals as a function time, 
where $p$ is the fractal dimension of the stochastic DCS,
showing that $M_p(t)$ is a conserved quantity in the long-time limit. The three distinct parallel lines for the
same probability $p$ reveal that the numerical
value of the conserved quantity $M_p(t)$,  like kinetic DCS, is different in each independent realization.
}
\label{fig3}
\end{figure}

We can make Eq. (\ref{eq:binary_frag}) describing the rules of the 
stochastic DCS problem if  we only choose 
\begin{equation} 
F(x,y) =1, 
\end{equation}
and at the same time replace the factor $2$ of the gain term again by $(1+p)$.
The master equation for the stochastic DCS then is 
\begin{equation}
\label{eq:sdcs} 
{{\partial c(x,t)}\over{\partial t}}= -xc(x,t)+(1+p)\int_x^\infty c(y,t)dy.
\end{equation}
The stochastic counterpart of the dyadic Cantor set is much simpler to solve analytically than the stochastic counterpart of the triadic Cantor set which was
 first proposed and solved analytically by Ben-Naim and Krapivsky \cite{ref.hassan2, ref.krapivsky}. 
However, we for the first time give an exact algorithm of the model and focus primarily
on verifying  the various analytical results  
by extensive numerical simulation.  
Like before we once again incorporate the definition of the $n$th moment in the rate
equation to obtain
\begin{equation}
{{dM_n(t)}\over{dt}}= -\Big [1-\frac{(1+p)}{(n+1)}\Big ]M_{n+1}(t).
\end{equation}
Following the same procedure as for the kinetic DCS problem, we obtain the asymptotic 
solution for the $n$th moment
\begin{equation}
\label{eq:moment_sdcs}
M_n(t)\sim t^{(n-p)z} \hspace{0.35cm} {\rm with}  \hspace{0.35cm}
z=-1,
\end{equation}
which implies that $p$ is the special value of $n$ for which $M_p(t)$ is a conserved quantity (see Fig (3)).

\begin{figure}
\includegraphics[width=5.50cm,height=8.5cm,clip=true,angle=-90]{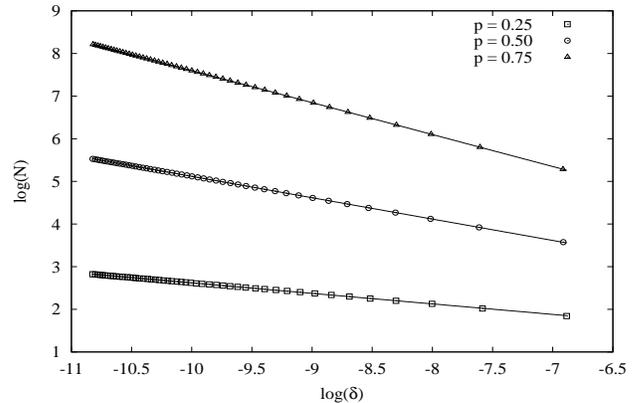}
\caption{${\rm Log}-\log$ plot of the interval number $N$ vs yard-stick size $\delta$ using simulation data for three different $p$
values and the same initial condition. The lines have slopes
$d_f=p$ depicting that $N(\delta)\sim \delta^{-d_f}$ and hence this is in perfect agreement with our analytical result given by
Eq. (\ref{ndelta}).
}
\label{fig4}
\end{figure}

Note that the solution for $M_n(t)$ given by Eq. (\ref{eq:moment_sdcs}) clearly suggests that the dynamics of the
stochastic DCS too governed by a conservation law namely the moment of 
order $n=p$ is a conserved quantity. 
The solution for $M_n(t)$ also suggests that the mean interval size $\delta(t)$ decays following the same power-law and 
the same exponent as in Eq. (\ref{eq:mean_krdcs}). Using it as the yard-stick we find that the number $N(\delta)$ 
needed to cover the resulting set scales as
\begin{equation}
\label{ndelta}
N(\delta) \sim \delta^{-d_f},
\end{equation}
with $d_f=p$  revealing that the index of the conserved quantity is again equal to the fractal dimension (see Fig. (4)).
Note that the fractal dimension $d_f=p$ is always less than ${{\ln (1+p)}\over{\ln 2}}$ regardless of the value of $p$.
Hence we can conclude that the fractal dimension of 
the stochastic fractal is always less than that of its recursive or kinetic counterpart. Once again we find that the moment 
of order equal
to fractal dimension $M_{d_f}$ is a conserved quantity. It seems quite like a rule than otherwise as we have found it to true
 also in the
case of its opposition phenomena namely in aggregation \cite{ref.fractalCDA, ref.fractalself_repl,
ref.hassanRogers1,ref.hassanRogers2,ref.hassan1,ref.hassan2}.

We shall now apply the Buckingham $\pi$-theorem to obtain a scaling solution for $c(x,t)$ as it provides a deeper insight into the problem 
\cite{ref.barenblatt}.
Note that according to Eq. (\ref{eq:sdcs}) the governed parameter $c$ depends on three parameters
$x$, $t$ and $p$ of which only the former two are dimensional. However,
the knowledge about the decay law for the mean interval size implies that one of the parameter, say $x$, can be expressed
in terms of $t$ since we find $\delta= t^{-1}$ must bear the dimension of the
interval size $x$. We therefore can define a dimensionless governing parameter
\begin{equation}
\xi ={{x}\over{t^{-1}}},
\end{equation}
and a dimensionless governed parameter
\begin{equation}
\label{eq:Pi_1}
\Pi={{c(x,t)}\over{t^\theta}}.
\end{equation}
The numerical value of the right side of the above equation remains the same even if the unit time $t$ is changed by
some factor, say $\mu$ for example, since the left hand side is a dimensionless quantity. It
means that the two parameters $x$ and $t$ must combine to form a dimensionless quantity 
$\xi=x/t^{-1}$ and the dimensionless parameter $\Pi$ only depends on $\xi$. In other words
we can write 
\begin{equation}
\label{eq:Pi_2}
{{c(x,t)}\over{t^\theta}}=\phi(x/t^{-1}),
\end{equation}
which leads to the following dynamic scaling form
\begin{equation}
\label{eq:scaling_form}
c(x,t)\sim t^\theta \phi(x/t^{-1}), 
\end{equation}
where the exponent $\theta$ is fixed by some conservation law and $\phi(\xi)$ is known as the scaling function.
Indeed, substituting Eq. (\ref{eq:scaling_form}) in the conservation law 
\begin{equation}
\int_0^\infty x^pc(x,t)dx={\rm const.},
\end{equation}
we immediately obtain that $\theta=1+p$.

We now substitute Eq. (\ref{eq:scaling_form}) in Eq. (\ref{eq:sdcs}) and find that it reduces the partial integro-differential equation
 into an ordinary integro-differential equation for the  scaling function $\phi(\xi)$ given by
\begin{equation}
\label{eq:scaling_1}
\xi{{d\phi(\xi)}\over{d\xi}}+\Big (\xi+(1+p)\Big )\phi(\xi)=(1+p)\int_\xi^\infty\phi(\eta)d\eta.
\end{equation}
This is much simpler to solve than solving Eq. (\ref{eq:sdcs}). To simplify it even further, we differentiate 
Eq. (\ref{eq:scaling_1}) with respect to $\xi$ getting
\begin{equation}
(-\xi){{d^2\phi(\xi)}\over{d(-\xi)^2}}+\Big ((2+p)-(-\xi) \Big ){{d\phi(\xi)}\over{d(-\xi)}}-(2+p)\phi(\xi)=0,
\end{equation}
This is exactly Kummer's confluent differential equation whose solution is given by
\begin{equation}
\phi(\xi)=~_1F_1(2+p;2+p;-\xi),
\end{equation}
where $_1F_1(a;c;z)$ is called the Kummer's confluent hypergeometric function \cite{ref.luke}. We find that 
in the limit $\xi\rightarrow \infty$ the solution for the function $\phi(\xi)$  assumes a simple form 
\begin{equation}
\phi(\xi)=~e^{-\xi}.
\end{equation}
Using it in Eq. (\ref{eq:scaling_form}) we find 
\begin{equation}
\label{eq:cxt_sdcs}
c(x,t)\sim t^{1+p}e^{-xt},
\end{equation}
which essentially implies that it exhibits dynamic scaling in the limit $ t \rightarrow \infty $ \cite{ref.family_Vicsek}. 
To verify it we have drawn a histogram of the distribution function 
$c(x,t)$ collecting data for fixed a time
in the log-linear scale and found a set of straight line since for a fixed time the interval size
 distribution function $c(x,t)$ should decay exponentially (see Fig. (5)).

\begin{figure}
\includegraphics[width=5.50cm,height=8.5cm,clip=true,angle=-90]{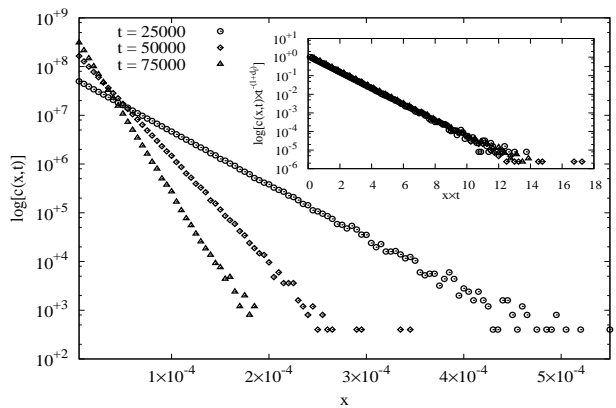}
\caption{Log-linear plot of the interval size distribution function $c(x,t)$ vs $x$ for the stochastic DCS 
using data at three different times. 
Inset shows the collapse of the same data when we plot  $t^{-(1+p)}c(x,t)$ vs $xt$ in the ${\rm Log}-{\rm linear}$ scale
which clearly verifies that the solution for $c(x,t)$ given by Eq. (\ref{eq:cxt_sdcs}).
}
\label{fig5}
\end{figure}

One of the key features of fractal is that it must be self-similar. 
In general, by self-similarity we mean that a suitably chosen part of an 
object represents the whole. The question is: What do we mean by self-similarity in stochastic fractal? Note that stochastic 
fractal such as stochastic DCS is not static rather 
it evolves with time. By self-similarity in such a case, we mean that it is similar with itself at different times. Note that the 
same system at different times
are similar if the numerical values of various dimensional governing parameters are different, however, the numerical values of 
the corresponding dimensionless quantities must coincide.
Indeed, in the case of stochastic DCS the numerical value of the governed parameter $c(x,t)$ for a given value of the governing 
parameter $x$ is different at different time which can easily be seen in Fig (5). 
However, the numerical value of the corresponding dimensionless governed parameter $c(x,t)/t^{1+p}$ and 
the dimensionless governing parameter $xt$ should coincide. 
 That is, all the distinct sets of curve of $c(x,t)$ as a function of $x$ at different times 
should collapse onto one single curve if we plot $c(x,t)/t^{1+p}$ as a function of $xt$.
This is exactly what we have shown in the inset of Fig. (5) and found that the data points of all the three distinct curves of Fig (5) merge superbly onto a 
single universal curve which is essentially the scaling function $\phi(x)$. 

\section{Self-similarity in kinetic DCS}

\begin{figure}
\includegraphics[width=5.50cm,height=8.5cm,clip=true,angle=-90]{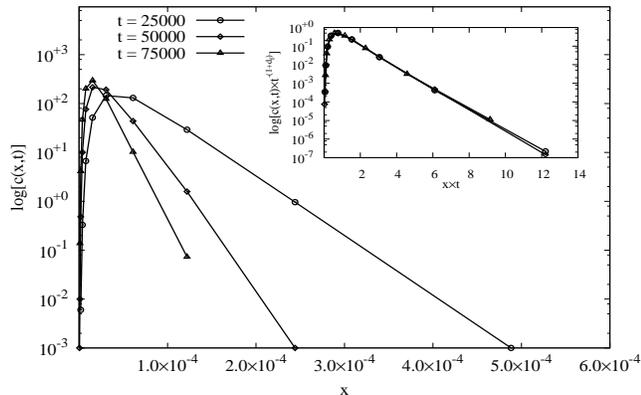}
\caption{${\rm Log}-{\rm linear}$ plots of the interval size distribution function $c(x,t)$ of the kinetic DCS is drawn as a function of $x$ 
for three different times. In inset we show that the three distincts curves for three different system sizes collapsed onto 
a single universal curve if we plot 
$t^{-(1+d_f)}c(x,t)$ vs $xt$. It implies that the system exhibits self-similarity with respect to time.
}
\label{fig 8}
\end{figure}

What about self-similarity in the kinetic dyadic Cantor set? Does it also exhibit dynamic scaling? A closer look at the solution 
$c(x,t)\sim t^{1+p}e^{-xt}$ for the stochastic DCS perhaps can guide us to write the solution of $c(x,t)$ for the kinetic DCS. To this end, we find it 
instructive to express $c(x,t)$ of stochastic DCS in terms of the mean interval size $\delta(t)$ and  the fractal dimension $d_f$
\begin{equation}
c(x,t)\sim \delta^{-(1+d_f)}\phi(x/\delta(t)),
\end{equation}
as we know $\delta\sim t^{-1}$ and $d_f=p$ for stochastic DCS. That is, substituting  $d_f=\ln (1+p)/\ln 2$ and $\delta\sim t^{-1}$ for kinetic DCS we can immediately
write the expected solution for it as
\begin{equation}
c(x,t)\sim t^{1+{{\ln (1+p)}\over{\ln 2}}}\phi(xt).
\end{equation}
To verify it, we plot first $c(x,t)$ vs $x$ for three different times (see Fig. (6)) and then plot the corresponding dimensionless quantities $c(x,t)/t^{1+{{\ln (1+p)}\over{\ln 2}}}$
vs $xt$. We once again find a superb data-collapse which is shown in the inset of the Fig (6) revealing that the fractal generated by kinetic DCS 
too exhibits a
dynamic scaling {\it vis-a-vis} self-similarity.

\section{Discussion and summary}

In this section we attempt to summarize and put all our findings into perspective. In this article we proposed a
dyadic Cantor set and then its kinetic and stochastic counterpart. In these models we incorporated probability, time and randomness 
in such a way that a logical flow can easily be appreciated. 
Firstly, We solved the DCS problem by using probabilistic argument and found that it emerges as a fractal of dimension 
$d_f=\ln (1+p)/\ln 2$.
Then we solved the kinetic and stochasticc DCS using the rate equation approach and found that the resulting systems
emerge as fractal of dimensions $d_f=\ln (1+p)/\ln 2$ and $d_f=p$ respectively. Our studies reveal that 
the inclusion of kinetics alone does not alter the value of the fractal dimension. However, inclusion of randomness 
in the generator of the stochastic DCS does change the value of the fractal dimension and found that it
is always smaller than that of the DCS or the kinetic DCS.

Besides giving exact analytical expressions for fractal dimensions we  have  also given solutions for the
interval size distribution function $c(x,t)$ for both kinetic and stochastic DCS problem 
alongside numerical simulations which fully support our solutions. In particular, we have shown that
the distribution function $c(x,t)$ evolves to a state where it exhibits dynamic scaling $c(x,t)\sim t^{1+p}e^{-xt}$. 
We used the idea of data-collapse to verify it numerically. For instance, we have drawn $c(x,t)$ vs $x$ using data 
for a fixed time and found a distinct set of curves for every different time of the process. However, 
these distinct curves collapsed onto a single universal curve when we 
expressed $c(x,t)$ in unit of $t^{1+p}$ and $x$ in unit of $t^{-1}$. It implies that the system is similar to itself 
at different time and hence we say that such systems exhibit self-similarity. It readily implies that the solution is 
independent of initial condition. 
We have shown that the self-similar properties in all three cases are accompanied by conservation laws
and interestingly the value of the fractal dimension $d_f$ coincide with
the index of the conserved quantity  $M_{d_f}$. The numerical value of the conserved quantity 
$M_{d_f}$, however, is found different at each realization. We have then checked if the $d_f$th moment is also a conserved 
in the dyadic  and traditional triadic Cantor set
and found that it is obeyed in both the cases as well.  
We think it is quite safe to argue that 
when systems evolve and yet preserve self-similarity then the dynamics of the system must be governed by conservation law.

To explain why the quantity $M_{d_f}$ always remains constant
we find it useful to consider a simple dimensional analysis.
According to Eq. (\ref{eq:cxt_sdcs}) the 
physical dimension of $c(x,t)$ is $[c]=L^{-(1+d_f)}$ since $[s(t)]=L$. 
 To know Why the quantity $M_{d_f}=\int_0^\infty x^{d_f}c(x,t)dx$ a conserved quantity we find it convenient
 to look into the physical dimension of its differential quantity 
$dM_{d_f}=x^{d_f}c(x,t)dx$. Using
the physical dimension $[x]=L$ and $[c(x,t)]=L^{-(1+d_f)}$ in the expression for $dM_{d_f}$, 
we immediately find that it bears no dimension and so does the quantity $M_{d_f}$. It implies that the numerical value of $M_{d_f}$ 
remains the same despite the fact that the system evolves with time.
On the other hand, the 
concentration $c(x,t)$ is defined as 
the number of particles per unit volume of embedding space ($V\sim L^d$ where $d=1$) per unit mass ($M$) and hence
$[c]=L^{-1}M^{-1}$  \cite{ref.rajesh}. Now applying the principle of equivalence we obtain
\begin{equation}
\label{eq:masslength}
M(L)\sim L^{d_f},
\end{equation}
which is often considered as one of the most commonly used benchmark for defining fractal.
An object whose mass-length relation satisfies Eq. (\ref{eq:masslength}) with typically a non-integer exponent is said to be
a fractal.

In summary, besides solving the model analytically, we also performed extensive  numerical simulation which fully support
all theoretical findings. Especially, the conditions under which scaling and fractals emerge
are found, the fractal dimensions of the three models are
given and the relationship between fractal dimension and 
conserved quantity is pointed out. Our findings complement the results found in the condensation-driven 
aggregation indicating
that these results are ubiquitous in the aggregation processes where mass conservation is violated. 
Besides, we show that the interval size distribution function in both variants of DCS exhibits 
dynamic scaling and we verify it numerically using the idea of data-collapse.
We hope this work will provide
useful insights and theoretical predictions for various physical, chemical and biological systems that emerge as fractal.
It would be instructive to analyze our model with other fragmentation
rates described by sum kernel $K(x,y) = x+y$ and product kernel $K(x,y) = xy$.  We propose to
address these issues in subsequent work.

 NIP would like to thank Dhaka University for awarding Bangabondhu Fellowship.

\end{document}